\documentclass[amsmath,amssymb,twocolumn,subscriptaddress,aps,showpacs,showkeys,preprintnumber]{revtex4}
\usepackage{bm,graphicx,subfigure,enumerate,latexsym}

\begin{document}
\title{Rheology of a supercooled polymer melt near an oscillating plate: an application of multiscale modeling}
\author{Shugo Yasuda}
\author{Ryoichi Yamamoto}
\affiliation{
Department of Chemical Engineering, 
Kyoto University, Kyoto 615-8510, Japan
and
CREST, Japan Science and Technology Agency, Kawaguchi 332-0012, Japan.
}

\begin{abstract}
The behavior of a supercooled polymer melt composed of short chains with ten beads near an oscillating plate are simulated by using a hybrid simulation of molecular dynamics (MD) and computational fluid dynamics (CFD).  In the method, the macroscopic dynamics are solved by using CFD, but, instead of using any constitutive equations, a local stress is calculated by using a non-equilibrium MD simulation associated at each lattice node in the CFD calculation.
It is seen that the local rheology of the melt varies considerably in a thin viscous diffusion layer that arises near an oscillating plate.
It is also found that the local rheology of the melt is divided into the three different regimes, i.e., the viscous fluid, viscoelastic liquid, and viscoelastic solid regimes, according to the local Deborah number $De$, which is defined with the Rouse or $\alpha$ relaxation time, $\tau_R$ or $\tau_\alpha$, and the angular frequency of the plate $\omega$ as $De^R$=$\omega\tau_R$ or $De^\alpha$=$\omega\tau_\alpha$.
The melt behaves as a viscous fluid when $De^R\lesssim 1$, and the crossover between the liquid-like and solid-like regime takes place around $De^\alpha\simeq 1$.
\end{abstract}

\maketitle
Due to the emergence of nano-science (or nano-technology), a great innovation of the functional materials of soft matters is expected recently.
An important property of soft matter is that there exit strong correlations between the nano-scale structure and kinetics and the macroscopic  behavior of the system.
In other words, the nano-scale structures and kinetics may affect the macroscopic physical properties, and the macroscopic behavior may also affect the microscopic physics.
This property of the coupling of the microscopic and macroscopic physics is very useful for functional materials, but it also makes the theoretical and experimental investigations of soft matter difficult.
The simulations may be useful for predicting the complex behaviors of soft matter. 
The difficulty in the simulation of soft matter is the coexistence of the macroscopic and microscopic scales involved. 
The macroscopic dynamics are usually solved by using the continuum equation, e.g., the Navier-Stokes equation, with a constitutive relation.
For soft matter, however, the constitutive relation is so complicated that no explicit formulas are obtained in general.
In contrast, the molecular dynamics simulations are performed without using any constitutive relations.
However, the time and space to be simulated are restricted to very microscopic scales.
In order to overcome this difficulty, we have recently developed a hybrid simulation of the molecular dynamics (MD) and the computational fluid dynamics (CFD) based on a local sampling strategy, in which the macroscopic dynamics are solved using a CFD scheme but, instead of using any constitutive equations, a local stress is calculated by using a non-equilibrium MD simulation associated with each lattice node of the CFD computation.\cite{art:08YY}
The basic idea of the hybrid simulation method was put forward earlier by Kevrekidis {\it et al.}\cite{art:03KGHKRT} and also by Ren and E.\cite{art:05RE}

In previous papers, the validity of the hybrid methods and their efficiencies are examined intensively for viscous fluids without memory effects. 
De {\it et al.} have recently proposed a new hybrid method, which is called the scale bridging method in their paper, that can correctly reproduce the memory effect of the polymeric liquid, and performed a simulation of a non-linear viscoelastic polymeric liquid between two parallel oscillating plates.\cite{art:06DFSKK}
They have also compared the results obtained by the scale bridging method with those obtained by a full MD simulation, thereby demonstrating validity of the method. 
In the present letter, we also model the behavior of a polymer melt between two parallel oscillating plates by using the hybrid method, but we focus on the complex rheology of a supercooled polymer melt in the viscous diffusion layer that arises near the fast oscillating plate. 
The boundary layer arises if the width between the plates is much larger than the thickness of viscous diffusion layer.
Note that, in Ref. \onlinecite{art:06DFSKK}, the thicknesses of viscous diffusion layers are estimated to be comparable to the widths of the plates\cite{art:05SKK}, thus the boundary layers are not clearly seen.

In the present problem, the macroscopic quantities are quite non-uniform, and two different characteristic length scales appear that must be resolved; one is that of a polymer chain and the other is that of a boundary layer arising near the oscillating plate. 
This problem constitutes an important application of multiscale modeling since it is quite difficult to solve this problem by using a full MD simulation because the length of the boundary layer is much larger than that of a polymer chain.
In the following, we briefly state the problem and outline the hybrid simulation method, and then discuss the numerical results. 
Finally we give a summary of our conclusions. 

\begin{figure}[tb]
\includegraphics[scale=1]{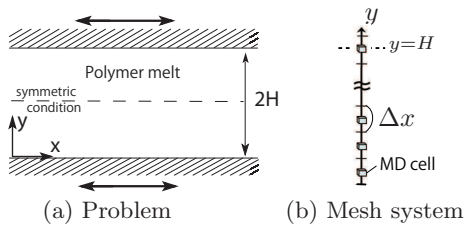}
\caption{
Schematics for the problem and mesh system.
}\label{f1}
\end{figure}

We consider a polymer melt with a uniform density $\rho_0$ and a temperature $T_0$ between two parallel plates (see Fig. \ref{f1}(a)). 
The upper- and lower-plate start to oscillate in opposite, parallel directions with a constant frequency $\omega_0$ at $t$=0. 
The polymer melt is composed of short chains with ten beads.
All of the bead particles interact with a truncated Lennard-Jones potential defined by\cite{art:90KG}
$U_{\rm LJ}(r)=
4\epsilon\left[
({\sigma}/{r})^{12}
-({\sigma}/{r})^{6}
\right]
+\epsilon$ 
for $r\le 2^{1/6}\sigma$, and 
0 for $r> 2^{1/6}\sigma$.
By using the repulsive part of the Lennard-Jones potential only, we may prevent spatial overlap of the particles.
Consecutive beads on each chain are connected by an anharmonic spring potential,
$U_{\rm F}(r)=-\frac{1}{2}k_c R_0^2 \ln
\left[
1-({r}/{R_0})^2
\right]$,
with $k_c$=30$\epsilon/\sigma^2$ and $R_0$=$1.5\sigma$.
The number density of the bead particles is fixed at $\rho_0/m$=1/$\sigma^3$, where $m$ is the mass of the bead particle.
With this number density the configuration of bead particles becomes severely jammed at a low temperature, resulting in the complicated non-Newtonian viscosity and long-time relaxation phenomena typically seen in glassy polymers.
In the present letter, we fix the temperature at $T$=0.2$\epsilon/k$, where $k$ is the Boltzmann constant.\cite{art:02YO}

We assume that the macroscopic quantities are uniform in the $x$- and $z$-directions, i.e., $\partial/\partial x$=$\partial/\partial z$=0.
Then the macroscopic velocity $v_\alpha$ is described by the equations, $\rho_0{\partial v_x}/{\partial t} = {\partial \sigma_{xy}}/{\partial y}$ and $v_y$=$v_z$=0, where $\sigma_{\alpha\beta}$ is the stress tensor.
Here and afterwards, the subscripts $\alpha$, $\beta$, and $\gamma$ represent the index in Cartesian coordinates, i.e., \{$\alpha$,$\beta$,$\gamma$\}=\{$x$,$y$,$z$\}.
We also assume the non-slip boundary condition at the oscillating plate, $v_x$=$v_0\cos \omega_0 t$ at $y$=0 (where $v_0$ is a constant amplitude of the oscillation velocity), and the symmetric condition at $y$=$H$ (i.e., $v_x$($y$=$H$+$\Delta x/2$)=$-v_x$($y$=$H$-$\Delta x/2$)).
If the frequency $\omega_0$ is large enough, the thin viscous boundary layer forms over the oscillating plate. 
The thickness of the layer is estimated, for fluids with a constant viscosity $\nu_0$, as $l_\nu\sim 7\sqrt{\nu_0/\omega_0}$.
Note that the thickness of the viscous layer $l_\nu$ is much smaller than the width between the plates, $\l_\nu \ll H$, but is usually much larger than the scale accessible to a full MD simulation, in which the characteristic length scale is the length of the polymer chain $l_{\rm p}$, $l_{\nu} \gg l_{\rm p}$.

The constitutive relation of the stress tensor is quite complicated\cite{book:87BAH,book:88L,book:97HP}; the temporal value of the stress tensor of a fluid element depends on the previous values of the velocity gradients of the fluid element.
The relation may be written in a functional form as,
\begin{equation}\label{eq1}
\sigma_{\alpha\beta}(t,x_\alpha)=F_{\alpha\beta}[\kappa_{\alpha\beta}(t',x_\alpha'(t'))],
\quad {\rm with} \quad t'\le t,
\end{equation}
where $\kappa_{\alpha\beta}$ is the velocity gradient, $\kappa_{\alpha\beta}=\partial v_\alpha/\partial x_\beta$, and $x'_\alpha(t')$ represents the path line along which a fluid element has been moving.
In the one-dimensional problem, however, we don't need to consider the path line of the convective fluid element since the macroscopic velocity is restricted to be only in the $x$-direction where the macroscopic quantities are assumed to be uniform.
Thus the stress tensor for the present problem may be written in a functional involving the local strain rate,
\begin{equation}\label{eq2}
\sigma_{\alpha\beta}(t,y)=F_{\alpha\beta}[\dot \gamma(t',y)],
\quad{\rm with}\quad t'\le t,
\end{equation}
where $\dot \gamma$ is the strain rate, $\dot \gamma$=$\partial v_x/\partial y$.
Note that, although Eq. (\ref{eq2}) becomes much simpler than Eq. (\ref{eq1}), the temporal value of the local stress still depends on the previous values of the local strain rate. Its dependence is quite complicated, especially for glassy materials (for which explicit formulas are unknown in general). 
In our hybrid simulation, instead of using any explicit formulas for the constitutive relation, the local stress is generated by the non-equilibrium MD simulation associated with each local point. 
See Fig. \ref{f1} (b).

We briefly explain our hybrid simulation method.
We improve the previous hybrid method\cite{art:08YY} with regard to the time-integration scheme of the CFD and the treatment of the sampling duration of the MD so that the memory effects are correctly reproduced.
The basic idea of the present method is the same as that of the scale bridging method proposed in Ref. \onlinecite{art:06DFSKK}.
The computational domain [0,$H$+$\Delta x/2$] is divided into thirty-two slits with a constant width $\Delta x$. 
We use a usual finite volume method with a staggered arrangement, where the velocity is computed at the mesh node and the stress is computed at the middle of each slit.
One hundred chains are confined in each cubic MD cell with a side length $l_{\rm MD}$=10$\sigma$.
As for the time-integration scheme, we use the simple explicit Euler method with a small time-step size $\Delta t$.
The local stress at each time step of the CFD is calculated  by performing a non-equilibrium MD simulation according to the local strain rate in each MD cell.
The techniques of the non-equilibrium MD simulation are the same as those in the previous paper\cite{art:08YY}; we use the Lees-Edwards sheared periodic boundary condition and a Gaussian iso-kinetic thermostat to keep a constant temperature.
In the present problem, however, we cannot assume a local equilibrium state at each time step of the CFD simulation since the relaxation time of the stress may become much longer than the time-step size of the CFD simulation (in which the macroscopic motions of the system should be resolved).
In the present simulations, the simple time-average of the temporal stresses of the MD (averaged over the duration of a time-step of the CFD simulation) are used as the stresses at each time step of the CFD calculation without ignoring the transient time necessary for the MD system to be in steady state. 
Thus, the ratio of the time-step size of the CFD calculation $\Delta t$ to the sampling duration of the MD simulation $t_{\rm MD}$ is $\Delta t$/$t_{\rm MD}$=1.
The final configuration of molecules obtained at each MD cell is memorized as the initial configuration for the MD cell at the next time step of the CFD.
Thus we trace all of temporal evolutions of the microscopic configurations with a microscopic time step so that the memory effects can be reproduced correctly. 
Note that, compared with a full MD simulation, we can achieve an efficient computation with regard to the spatial integration by using MD cells that are smaller than the slit size used in the CFD simulation. 
The efficiency of the performance of our hybrid simulation is represented by the ratio of the slit size used in the CFD model $\Delta x$ to the cell size of the MD simulation $l_{\rm MD}$, $\Delta x$/$l_{\rm MD}$. Hereafter, the quantities normalized by the units of length $\sigma$ and time $\tau_0$=$\sqrt{m\sigma^2/\epsilon}$ are denoted with a superscript ``*''. 
In the following simulations, we fix the time-step size of the CFD simulation $\Delta t$, sampling duration of the MD simulation $t_{\rm MD}$ and time-step size of the MD simulation $\Delta \tau$ as $\Delta t^*$=$t_{\rm MD}^*$=1 and $\Delta \tau^*$=0.001, respectively. 
Thus, one thousand MD steps are performed in each MD cell at each time step of the CFD computation.

\begin{figure}[tb]
\includegraphics[scale=1]{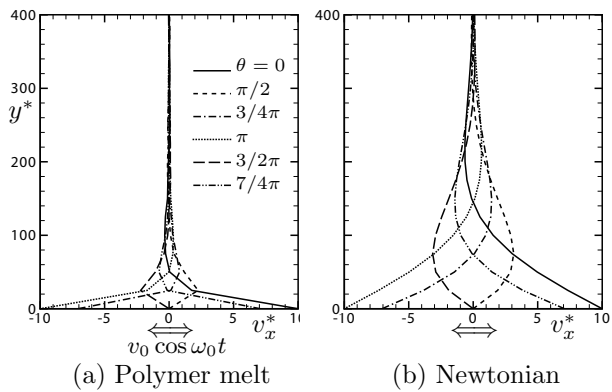}
\caption{
The velocity profiles at $\omega_0 t$=240$\pi$+$\theta$, $\theta/\pi$=0, 1/2, 3/4, 1, 3/2, and 7/4, for $\omega_0^*$=$2\pi/256$. (a) The result for the polymer melt and (b) that for a Newtonian fluid. The vertical axis represents the height $y$ and the horizontal axis the velocity $v_x$.
}\label{f2}
\end{figure}

We perform the hybrid simulations for two cases: Case A, in which $\omega_0^*$=$2\pi$/256 and $H^*$=787.5, and Case B, in which $\omega_0^*$=$2\pi$/1024 and $H^*$=1575.
The amplitude of the oscillation velocity is fixed as $v_0^*$=10 in both cases. 
The widths of the slits are $\Delta x^*$=25 for Case A and $\Delta x^*$=50 for Case B, and the ratios of the mesh size of the CFD $\Delta x$ to the cell size of the MD $l_{\rm MD}$ are $\Delta x$/$l_{\rm MD}$=2.5 for Case A and 5 for Case B. 
Figure \ref{f2} shows the instantaneous velocity profiles over a period for (a) Case A and (b) the Newtonian fluid with a constant viscosity $\nu_0^*$=$53$, which corresponds to the dynamical viscosity of the model polymer melt for $\omega_0$=$2\pi$/256.
The dynamical viscosity is calculated via $G_2^*/\omega$ (where $G_2$ is the loss modulus defined below).
It is seen that the boundary layer of the melt is much thinner than that of the Newtonian fluid.
This is caused by the shear thinning; the local loss modulus near the boundary is, as we see below, much smaller than that for the linear regime, thus the thickness of the viscous diffusion layer becomes thinner than that for the Newtonian fluid.
We also measure the local viscoelastic properties in terms of the local storage modulus $G_1(y)$ and loss modulus $G_2(y)$.
It should be noted that the local macroscopic quantities oscillate with a different phase retardation at each different point.
The local moduli are calculated in the following way:
The discrete Fourier transforms of the temporal evolutions of the strain $\gamma$, $\gamma(t,y)=\int_0^t\dot\gamma(t',y)dt'$, and shear stress $\sigma_{xy}$ during the steady oscillation states are performed.
The discrete Fourier transformations  are written as $\hat g_k^l$ = $\frac{1}{N+1}\sum_{n=0}^N g_n^l \exp({{\rm i}2\pi n k/(N+1)})$, with $g^l_n$=$g(n\Delta t,l\Delta x)$ ($n$=0,...,$N$ and $l$=0,...,32) , where $g$ represents the strain or shear stress (e.g., $g$=$\gamma$ or $\sigma_{xy}$).
By using the Fourier coefficients for the mode of the oscillating plate $k_0$, $k_0$=1+($\omega_0/2\pi$)$N$, the time evolution of the local strain at $y$=$y^l$ can be expressed as a cosine function,
\begin{equation}\label{eq8}
\gamma^l(t) = \gamma^l_0\cos( \omega_0 t + \delta^l),
\end{equation}
with $\gamma^l_0$ = $\sqrt{(\hat {\gamma'}_{k0}^l)^2+(\hat {\gamma''}_{k0}^l)^2}$ and $\delta^l$ = $\tan^{-1}(\hat {\gamma''}_{k0}^l/\hat {\gamma'}_{k0}^l)$.
Hereafter the superscripts ``$\prime$'' and ``$\prime\prime$'' indicate the real and imaginary part of the discrete Fourier coefficients, respectively.
The time evolution of the local shear stress can also be expressed as
\begin{equation}
\sigma_{xy}^l(t) = \sigma_1^l\cos(\omega_0 t + \delta^l) - \sigma_2\sin(\omega_0 t + \delta^l),
\end{equation}
with $\sigma_1^l$ = $\hat {\sigma'}_{k_0}^l\cos \delta^l + \hat {\sigma''}_{k_0}^l\sin \delta^l$ and $\sigma_2^l$ = $\hat {\sigma''}_{k_0}^l\cos \delta^l - \hat {\sigma'}_{k_0}^l\sin \delta^l$.
Thus, the local storage modulus $G_1$ and loss modulus $G_2$ are written, respectively, as $G_1(y^l)$=$\sigma_1^l/\gamma_0^l$ and $G_2(y^l)$=$\sigma_2^l/\gamma_0^l$.
\begin{figure}[tb]
\includegraphics[scale=1]{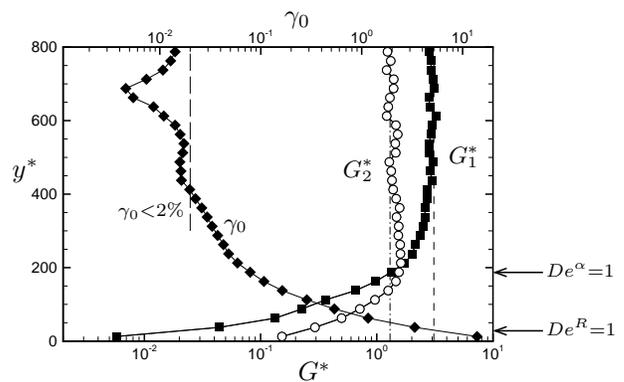}
\caption{
The spatial variations of the local moduli $G_1$ and $G_2$ (lower axis) and the amplitude of the local strain $\gamma_0$ (upper axis) for Case A. The dashed and dash-dotted lines show the storage modulus and loss modulus for the linear regime, respectively. The long-dashed line represents $\gamma_0$=2$\%$. The left arrows on the right-side vertical axis show the positions where the local Deborah numbers, shown in Fig. 5, are equal to unity. 
}\label{f3}
\end{figure}%
\begin{figure}[tb]
\includegraphics[scale=1]{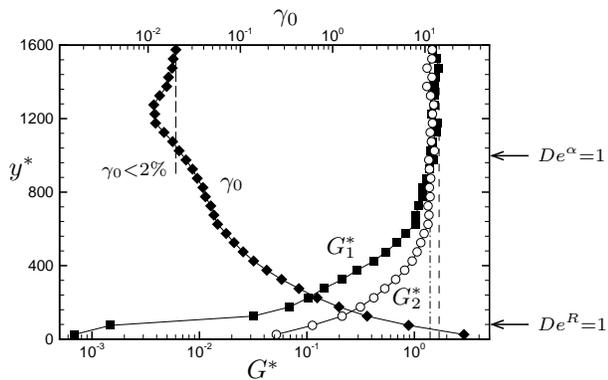}
\caption{
The spatial variations of the local moduli $G_1$ and $G_2$ (lower axis) and the amplitude of the local strain $\gamma_0$ (upper axis) for Case B. See also the caption for Fig. 3.
}\label{f4}
\end{figure}%
\begin{figure}[tb]
\includegraphics[scale=1]{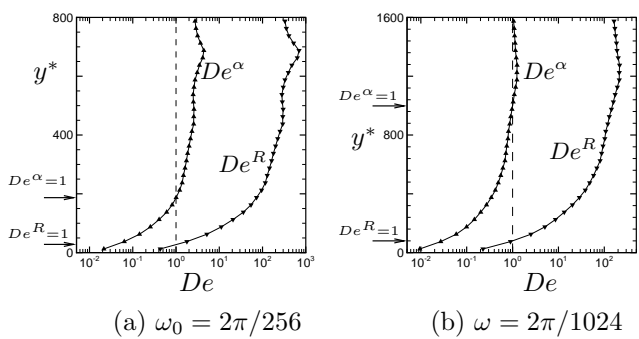}
\caption{
The local Deborah numbers (defined via the Rouse relaxation time $\tau_R$ and the $\alpha$ relaxation time $\tau^\alpha$) $De^R=\omega_0\tau_R$ and $De^\alpha=\omega\tau^\alpha$. (a) Case A. (b) Case B. The right arrows on the $y$-axis show the position where the Deborah numbers are equal to unity.
}\label{f5}
\end{figure}%

Figure \ref{f3} shows the spatial variations of the local storage modulus and loss modulus and the amplitude of the local strain for Case A.
The shear thinning is seen near the oscillating plate; the strain $\gamma_0$ (or the strain rate $\dot\gamma_0$=$\omega_0\gamma_0$) is quite large near the oscillating plate and both moduli are much smaller than those for the linear regimes.
In the close vicinity of the oscillating plate, the storage modulus $G_1$ is much smaller than the loss modulus $G_2$, $G_1 \ll G_2$. 
Hence, the melt behaves as a viscous fluid. 
The storage modulus rapidly grows with the distance from the oscillating plate, and the viscoelastic crossover occurs at $y^*\sim 200$. 
Both moduli attain their linear values, which are shown as dashed and dot-dashed lines in Fig. \ref{f3}, for distance that is far from the oscillating plate where the local strains are less than about two percent. 
The overall features are also consistent with Case B, although the crossover is not as clear as that in Case A (See Fig. \ref{f4}). 
Thus, the local rheology of the melt can be divided into three regimes, i.e., the viscous fluid, viscoelastic liquid, and viscoelastic solid regimes. 
These regimes may be also characterized by the two ``local'' Deborah numbers. 
One is defined by the local Rouse relaxation time $\tau_R$ of the melt and the angular frequency of the plate $\omega_0$, $De^R$=$\omega_0\tau_R$, and the other is defined by the local $\alpha$ relaxation time $\tau_\alpha$ and the angular frequency $\omega_0$, $De^\alpha$=$\omega_0\tau_\alpha$. 
Note that the local Rouse and $\alpha$ relaxation times vary according to the local strain rate $\dot \gamma$, $\tau=\tau(\dot\gamma)$. 
Figure \ref{f5} shows the spatial variation of the local Deborah number $De^R$ and $De^\alpha$, where the local relaxation times $\tau_R$ and $\tau_\alpha$ are calculated by substituting the values of $\dot \gamma_0^l$, which are obtained via Eq. (\ref{eq8}), into the fitting functions for the relaxation times for the simple shear flows obtained in Ref. \onlinecite{art:02YO}.
In Figs. \ref{f3} and \ref{f4}, the positions at which the local Deborah numbers become equal to unity are shown by the left arrows. 
It is seen that the melt behaves as a viscous fluid, $G_2\gg G_1$, for $De^R\lesssim 1$, while the viscoelastic properties become pronounced for  $De^R\gtrsim 1$. 
This is consistent with the rheology diagram for a model polymer melt obtained in Ref. \onlinecite{art:06VB}.
It is also seen that the crossovers of the liquid-like regime, $G_2>G_1$, and the solid-like regime, $G_1>G_2$, take place at $De^\alpha\sim 1$.

In summary, we have numerically analyzed the behaviors of a supercooled polymer melt near an oscillating plate by using the hybrid simulation of MD and CFD  based on the local sampling strategy. 
In our simulation, the memories of the molecular configurations of the fluid elements are correctly traced at the microscopic level. 
The efficiencies of our hybrid simulations are represented by the ratios of the mesh size of the CFD simulation, $\Delta x$ to the cell size of the MD simulation, $l_{\rm MD}$.  
These ratios are $\Delta x/l_{\rm MD}$=2.5 for Case A and 5 for Case B. It is found that the local rheology of the melt varies considerably in a viscous boundary layer near an oscillating plate, so that three different regimes, i.e., the viscous fluid, viscoelastic liquid, and viscoelastic sold regimes, form over the oscillating plate. 
It is also found, in the viscous fluid regime, that the local Deborah number defined via the Rouse relaxation time and the angular frequency of the plate is less than about unity, $De^R\lesssim 1$. 
The crossover between the liquid-like and solid-like regimes takes place around the position where the local Deborah number defined via the $\alpha$ relaxation time and the angular frequency is equal to unity, $De^\alpha\sim 1$.
\bibliographystyle{apsrev}
%

\end{document}